\documentclass[a4paper,fleqn,usenatbib]{mnras}

\usepackage{newtxtext,newtxmath}

\usepackage[T1]{fontenc}
\usepackage{ae,aecompl}
\usepackage{hyperref}

\usepackage{lineno}

\usepackage{graphicx}	
\usepackage{amsmath}	
\usepackage{amssymb}	
\usepackage{mathtools,amsmath}

\title[Gamma rays from reaccelerated particles at SNRs]{Gamma rays from reaccelerated particles at supernova remnant shocks}

\author[P. Cristofari \& P. Blasi]{
P. Cristofari$^{1,2}$ \& P.~Blasi$^{1,2}$\thanks{E-mail: pierre.cristofari@gssi.it}
\\
$^{1}$Gran Sasso Science Institute, via F. Crispi 7--67100, L'Aquila, Italy \\
$^{2}$INFN/Laboratori Nazionali del Gran Sasso, via G. Acitelli 22, Assergi (AQ), Italy
}
\date{Accepted XXX. Received YYY; in original form ZZZ}

\pubyear{2019}
\begin{document}
\label{firstpage}
\pagerange{\pageref{firstpage}--\pageref{lastpage}}
\maketitle

\begin{abstract}
Diffusive shock acceleration is considered as the main mechanism for particle energization in supernova remnants, as well as in other classes of sources. The existence of some remnants that show a bilateral morphology in the X-rays and gamma rays suggests that this process occurs with an efficiency that depends upon the inclination angle between the shock normal and the large scale magnetic field in which the shock propagates. This interpretation is additionally  supported by recent particle-in-cell simulations that show how ions are not injected if the shock is more oblique than $\sim 45^{o}$. These shocks provide an excellent test bench for the process of reacceleration at the same shock: non-thermal seed particles that are reached by the shock front are automatically injected and accelerated. This process was recently discussed as a possible reason for some anomalous behaviour of the spectra of secondary cosmic ray nuclei. Here we discuss how gamma--ray observations of selected supernova remnants can provide us with precious information about this process and lead us to a better assessment of particle diffusive shock reacceleration for other observables in cosmic ray physics. 

\end{abstract}



\section{Introduction}
Supernova remnants (SNRs) are of crucial importance for cosmic ray (CR) physics, as they are often thought to be major contributors to the Galactic CR spectrum up to the $\textit{knee}$, i.e. $\sim 1-3$ PeV. Several arguments support this idea~\citep[see e.g.][for reviews on this topic]{drury2012,blasi2013,amato2014}. The fact that SNRs have been observed in the gamma--ray domain is, by itself, an indisputable evidence that efficient particle acceleration does take place~\citep{FermiSNR,HESSSNR}, although it should be stressed that many of the observed gamma rays  from  SNRs can be interpreted in terms of radiating electrons. Moreover, at present there is no evidence of SNRs accelerating particles up to the energy of the knee. Whether this is a problem or not for the SNR paradigm depends on the stage of the SNR evolution when acceleration of PeV particles is expected to take place~\citep[see e.g.][]{cristofari2018}. In turn this depends on the type of supernova explosion and on the growth rate of CR induced instabilities leading to enhanced particle scattering (see for instance the work of \citet{Schure,Cardillo}).

Diffusive shock acceleration (DSA)~\citep{axford1977,krymskii1977,bell1978,blandford1978}, and its refined non--linear versions~\citep{berezhko1999,malkov2001,amato2006}, have helped describing the efficient acceleration of particles at SNR shocks, and the subsequent production of gamma rays, while studies based on hybrid and particle-in-cell (PIC) simulations have provided unprecedented insights into the microphysics of DSA \cite[]{2014ApJ...794...47C,2014ApJ...794...46C,caprioli2014}). There are two important ingredients of the theory, necessary to understand the acceleration to very high energy and to make sense of the multi-frequency spectrum of individual SNRs: 1) dependence of the injection upon the inclination angle between the shock normal and the local large scale magnetic field, and 2) the amplification of magnetic field as due to non-resonant streaming instability \cite[]{bell2004}. Recent hybrid simulations have provided a crucial improvement in our understanding of these two ingredients: \cite{caprioli2014} showed that the injection of thermal ions depends dramatically on the shock inclination, so that injection practically shuts off for inclinations larger than $\sim 45^{o}$. In turn, magnetic amplification cannot be triggered at such highly oblique shocks due to the lack of current in the form of escaping particles in the upstream direction. 
These two pieces of information hint at a possible reason for the existence of SNRs with a bilateral morphology in the X-ray band: non thermal X-ray radiation is the result of synchrotron emission by high--energy electrons in the amplified magnetic field. In the regions where the shock is oblique with respect to the shock normal, particles are not accelerated and magnetic field is not significantly amplified, hence the X-ray emission is also suppressed. A prototypical instance of such phenomenon is SN1006. 

The problem of particle injection is tightly connected with the microphysics of the formation of a collisionless shock and is related to the question of how many particles can actually cross the shock structure and complete one or a few cycles of Fermi acceleration, rather than being thermalized and advected downstream. On the other hand, if non--thermal seed particles were placed upstream of the shock with an energy that is large enough to cross the shock, these particles could be energized at the shock independently of the microphysical processes that give rise to the shock itself. This process is known as diffusive shock reacceleration (DSRA) and was already discussed in the pioneering work by \cite{bell1978,bell1978a}. All particles that happen to fill the upstream region of a SNR shock get energized. This phenomenon has recently been investigated \citep{blasi2017,bresci2019} in connection with the non trivial behaviour of the CR secondary-to-primary ratios as measured by AMS-02 \cite[]{ams}. Secondary nuclei, such as boron, that propagate through the Galaxy, have a finite chance of encountering a region where a SNR shock is present. Those nuclei are re-energized by the shock and acquire a harder spectrum, that eventually becomes important at sufficiently high energy and produces a trend in the B/C ratio that is different from the one expected in the standard model. Several questions arise from these considerations: what is the maximum energy of these reaccelerated particles? What are the shocks that are most important for DSRA (young and fast or old and slower)? Does DSRA lead to testable predictions in gamma rays or other wavelengths? What happens to Galactic CR protons (or nuclei) and electrons when they are reaccelerated at a SNR shock?

An important contribution to answering some of these questions may come from the search for direct evidence of DSRA in the gamma--ray spectra of some selected SNRs. In fact, gamma rays can either be produced through pp scattering of high--energy CR protons with ambient gas or through inverse Compton scattering (ICS) of high--energy electrons off 
optical, CMB, infra--red background photons. Typically the former mechanism is dominant in dense environments, while the latter is ubiquitous provided there are accelerated electrons. For the reasons described below, DSRA may be easier to spot in SNR shocks expanding in low density media and the gamma--ray emission is expected to be dominated by ICS of reaccelerated electrons, while reaccelerated protons have typically little effect on the emission. 

It is worth keeping in mind that although DSRA does not require specific recipes for injection, it does require the presence of scattering agents upstream of the shock. Such waves can be pre-existing or might even be self-generated, if some conditions get satisfied \cite[]{caprioli2017}.



Here we selected three cases of SNRs expanding in a dilute interstellar medium (ISM), for which a gamma--ray emission has been measured in the GeV and TeV energy range. We calculated the spectra of reaccelerated electrons and protons at the shocks of these SNRs, assuming that the seed particles (far upstream) have the same spectrum as measured in the local interstellar medium. The three SNRs we selected are SN~1006, RX J1713--3946 and RX J0852.0--4622 (Vela Jr) that satisfy the conditions listed above. We find that the gamma--ray emission contributed by ICS of reaccelerated electrons is sufficient to describe the observations, leaving little room for other components. For the case of SN1006, given the bilateral morphology, we derive specific limits of scattering properties in the regions where the magnetic field is expected to be quasi-parallel to the shock normal and highly oblique. 


The article is structured as follows: in Sec.~\ref{sec:spectrum} we describe the approach followed to calculate the spectrum of particles from DSRA at SNR shocks. In Sec.~\ref{sec:SN1006} we illustrate the calculation of the gamma--ray emission from SN 1006 as due to reaccelerated particles. In Sec.~\ref{sec:others} we comment on the case of two other SNRs, RX J1713--3946 and RX J0852.0--4622 (Vela Jr) with properties somewhat similar to SN 1006, and illustrate how future gamma--ray observations may help identifying the contribution of electrons from DSRA. We conclude in Sec.~\ref{sec:concl}.
  
 \section{Spectrum of reaccelerated particles} 
\label{sec:spectrum}

The energization of seed suprathermal particles at a shock was originally discussed in the pioneering articles by \cite{bell1978,bell1978a}. A non--linear theory of DSRA was developed by \cite{blasi2004}, that showed that under certain conditions the reacceleration of seed particles may modify the shock structure due to the pressure in the form of non-thermal particles ahead of the shock. The relative importance of the reaccelerated particles with respect to the freshly accelerated particles from the thermal pool depends on many factors: the orientation of the local magnetic field, the density of the gas and, clearly, the density in the form of seed suprathermal particles upstream. 

The description of the phenomenon of DSRA can be easily obtained by using the transport equation. Here we adopt the formalism of \cite{blasi2004}, although limited to the test particle case: a plane shock is considered with an axis $x$ oriented from upstream ($-\infty$) to downstream ($+\infty$), and the shock is assumed to be located at $x=0$. Indices 1 and 2 refer to quantities in the upstream and downstream region respectively.  The  stationary one-dimensional transport equation in the shock reference frame is:  
\begin{equation}
\label{eq:transport}
\frac{\partial }{\partial x}\left[D \frac{\partial f}{\partial x} \right] - u \frac{\partial f}{\partial x} + \frac{1}{3} \left( \frac{du }{dx} \right) p \frac{\partial f}{\partial p} + Q(x,p)=0 ,
\end{equation}
where $D(x,p)$ is the diffusion coefficient, $f(x,p)$ is the particle distribution in phase space, and $u(x)$ is the velocity of the plasma. $Q$ is the injection term with the usual form: 
\begin{equation}
 Q(x,p)= \frac{\eta n_1 u_1}{4 \pi p_{\rm inj}^2} \delta(x) \delta(p-p_{\rm inj}),
\end{equation}
where for simplicity it is assumed that injection only takes place at the shock surface and is strongly peaked around some momentum $p_{inj}$.  $n_1$ and $u_1$ are the density and velocity in the upstream region. 
The presence of seed particles is taken into account by imposing the boundary condition $f(x=-\infty,p)=f_{\infty}(p)$, where $f_{\infty}(p)$ is the distribution function (at upstream infinity) of the seeds to be reaccelerated \cite[]{bell1978,bell1978a}. This situation is supposed to mimic the expansion of a shock in the interstellar medium, where the background sea of CRs is present. The spectrum at the shock $f_0(p)=f(x=0,p)$ is obtained by integrating the transport equation from $-\infty$ to $0^-$, from $0^-$ to $0^+$ and assuming homogeneity downstream~\citep[see a detailed derivation in][]{blasi2004,blasi2017}. In the absence of non--linear effects, the integration easily leads to:
\begin{equation}
  \label{eq:f0}
  \begin{multlined}
f_0(p)= f_0^{\rm inj}(p) + f_0^{\rm seed}(p) \\
= \frac{s \eta n_1}{4 \pi p_{\rm inj}^3} \left( \frac{p}{p_{\rm inj}}\right)^{-s} + s \int_{p_0}^p \frac{\text{d}p' }{p'} 
\left( \frac{p'}{p} \right)^s f_{\infty}(p'),
 \end{multlined}
\end{equation}
where $s=3 r/(r-1)$, with $r=u_1/u_2$ the compression factor at the shock. The expression given in Eq.~\ref{eq:f0} introduces a minimum momentum $p_0$ for the reaccelerated particles, different from $p_{\rm inj}$. As discussed by~\citet{blasi2017}, the value adopted for $p_0$ is not crucial if the seeds are Galactic CRs. In fact, at momenta lower than $\sim 1$ GeV/c both the electron and proton spectra are harder than $p^{-4}$, as a result of energy losses and perhaps advection. As discussed below, this condition is sufficient to make the integral in the second term of Eq. \ref{eq:f0} to be dominated by momenta $p\sim mc$ for low momenta and by momenta larger than $mc$ for large momenta, but never by momenta around $p_{0}$. 

It is easy to show that, if the spectrum of seeds is steeper than $\sim p^{-s}$ the contribution of the reaccelerated particles approaches $\sim p^{-s}$. In practice this condition is met if we consider a strong shock $r\sim 4$ ($s \sim 4$) and the reacceleration of high--energy ($p>1$ GeV/c) Galactic CRs, where the spectrum of seeds $f_{\infty}$ is steeper than $\sim p^{-4}$, leading to a reaccelerated spectrum approaching $p^{-4}$. Remarkably, DSRA does not require any $\textit{efficiency}$ parameter  to be specified: in DSRA all particles are re-energized and the total energy transferred from the shock to the particles depends only upon the initial seed spectrum and the compression factor of the shock. In the typical cases that we will consider in this work,  our calculations lead to an energy in the final spectrum $\sim 30-70$ times larger than the one in the initial seeds. 

\begin{figure}
\includegraphics[width=.5\textwidth]{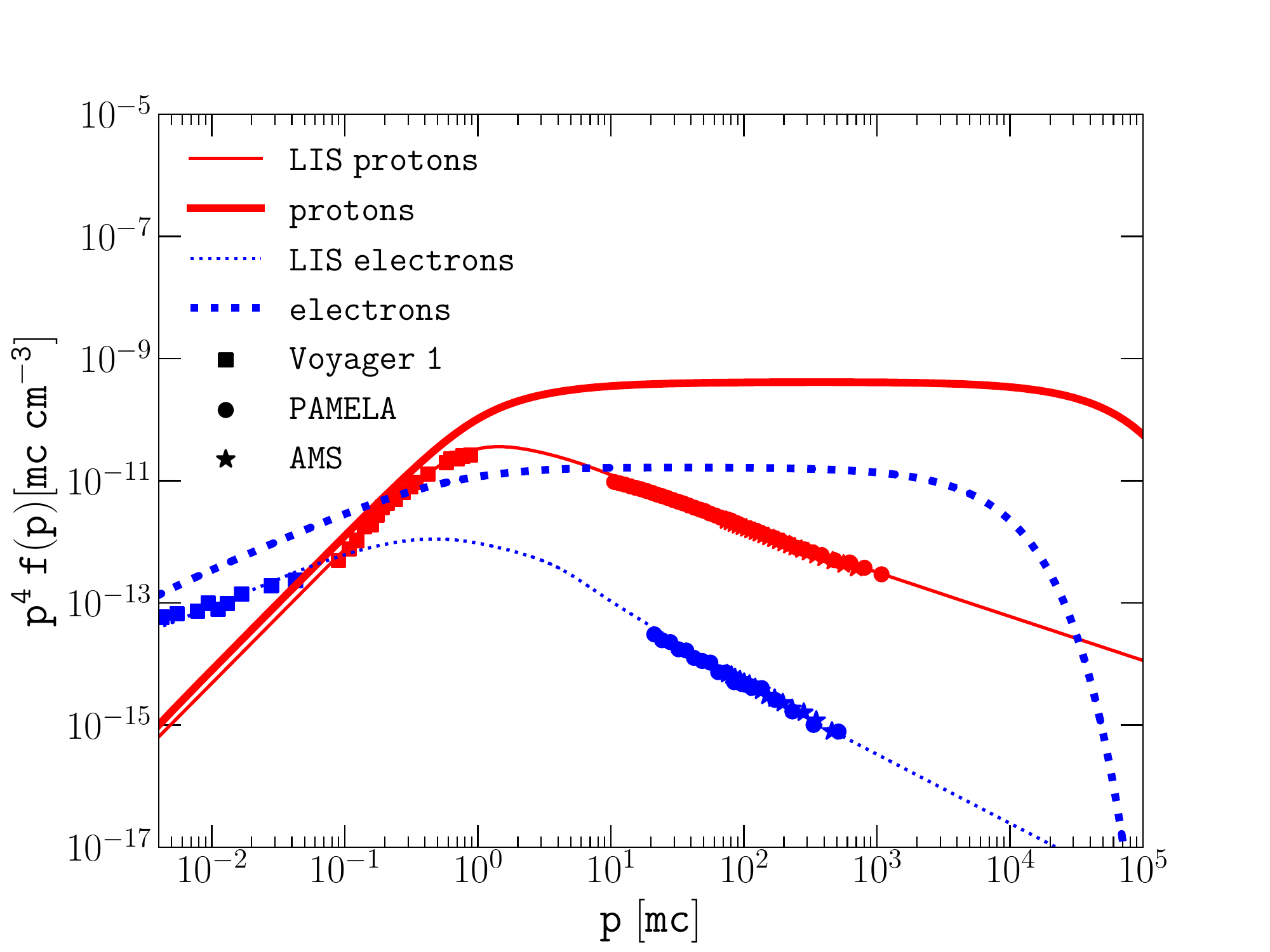}
\caption{Reaccelerated spectra (thick lines)  from seed protons (red solid lines) and electrons (blue dotted lines), following Eq.~\ref{eq:f0}. LIS  (thin lines) are taken from~\citep{bisschoff2019}. A cut--off is introduced in the electron spectrum in order to account for synchrotron losses as in Eq.~\ref{eq:tacc_min}. The data from Voyager (squares)~\citep{cummings2016}, PAMELA (dots)~\citep{adriani2011,adriani2011b} and AMS--02 (stars)~\citep{aguilar2015,aguilar2019} are shown. }
\label{fig:particles}
\end{figure}

 In Fig.~\ref{fig:particles}, we show the spectrum of Galactic protons $f_{\infty}^{p}$ (red thin solid line) and electrons $f_{\infty}^{e}$ (blue thin dotted line) assumed to be the same as in the local interstellar spectrum (LIS) and parametrized as proposed for protons and electrons by \citet{bisschoff2019}. These parametrizations describe well the Voyager I \cite[]{cummings2016} and PAMELA data \cite[]{adriani2011,adriani2011b}, as shown in Fig.~\ref{fig:particles}. We extrapolate these parametrizations to the energy range above 100 GeV, and in order to fit the AMS--02 data \cite[]{aguilar2015,aguilar2019}, we introduce a hardening in the proton and electron spectra, $\propto p^{0.1}$ and $\propto p^{0.2}$ respectively, at 300 GeV for protons and 100 GeV for electrons. It is worth stressing however that the details of such an extrapolation have no practical impact on the results that will follow.

 The spectra of reaccelerated protons and electrons are shown as thick lines  in Fig.~\ref{fig:particles}, and one can see that they both approach $\sim p^{-4}$ above $\gtrsim10$ GeV, as expected for a strong shock. No contribution from freshly accelerated protons and electrons was included here. One can immediately see that the net effect of DSRA is to transfer energy from low to high energies and harden the spectra when possible. In Fig.~\ref{fig:particles} we did not put much care in addressing the issue of the maximum energy of the reaccelerated particles, because the main purpose was to show the spectral shape. We simply assumed that the spectrum of electrons from DSRA is cutoff at $p_{\rm max}^{e}=10^4$ GeV and that the spectrum of protons extends to $p_{\rm max}^{p}=10^5$ GeV. 

On the other hand, we will see below that the issue of the maximum energy of electrons and protons from DSRA is central to the question of whether these particles produce observable signatures. Rather than embarking upon the investigation of the instability  growth induced by the current of reaccelerated particles, here we consider a few phenomenological situations that are supposed to bracket physically meaningful cases. Moreover we focus our attention on reaccelerated electrons, in that protons do not have a particularly interesting impact on observations, as we discuss below. We will illustrate the implications of the different cases for three SNRs, SN 1006, RX J1713--3946 and Vela Jr. in Secs.~\ref{sec:SN1006} and ~\ref{sec:others}. 

If there is enough magnetic field (perhaps generated by freshly accelerated protons) that the acceleration of electrons is limited by energy losses, and the diffusion coefficient is Bohm--like, then one can show that the spectrum of accelerated particles is cutoff with a shape $\propto \text{exp} \left[ -(p/p_{\rm max}^{e})^2\right]$. In the case of Kolmogorov--like diffusion coefficient, a cutoff with a shape $\propto \text{exp} \left[ -(p/p_{\rm max}^{e})\right]$ can be assumed~\cite[]{zirakashviliaharonian2007,blasi2010}.

The maximum momentum of reaccelerated electrons is estimated by equating the acceleration time to the minimum between the synchrotron loss time and the age of the SNR: 
\begin{equation}
\tau_{\rm acc} = \text{min} (\tau_{\rm sync}, t_{\rm age}).
\label{eq:tacc_min}
\end{equation}
Here the acceleration time is calculated as \cite[]{Drury}: 
\begin{equation}
\tau_{\rm acc}= \frac{3}{u_1-u_2} \int_{0}^{p}  \frac{\text{d}p'}{p'} \left( \frac{D_1(p')}{u_1} + \frac{D_2(p')}{u_2}\right).
\end{equation}

The uncertainty on the structure and strength of the magnetic field, and thus on the diffusion coefficient upstream and downstream of the shock, translates into an uncertainty in the maximum momentum of reaccelerated electrons (or even freshly accelerated electrons). 

\begin{figure}
\includegraphics[width=.5\textwidth]{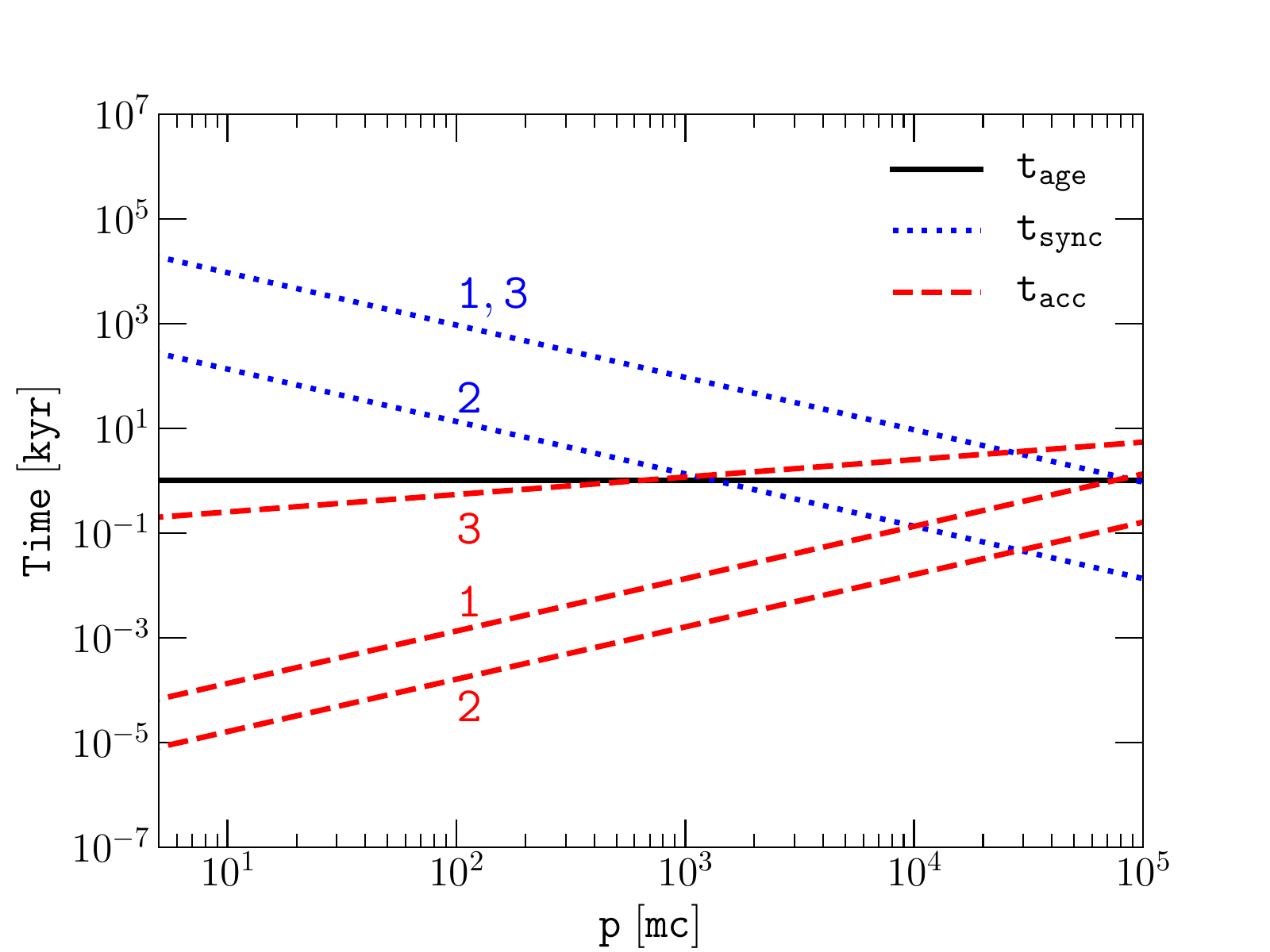}
\caption{Acceleration times $\tau_{\rm acc}$ (red dashed lines), synchrotron loss time $\tau_{\rm sync}$ (blue dotted lines) and $t_{\rm age}$ (black solid line) of SN 1006 as a function of momentum, for the three diffusion regime considered.}
\label{fig:times}
\end{figure}

For the sake of illustration of our results for a wide range of situations, here we consider three regimes of diffusion: (1) Bohm regime with $B_2 = r B_1$ and $B_1=3$ $\mu$G,  (2) Bohm regime with $B_2=100$ $\mu$G and (3) Kolmogorov regime where the diffusion coefficient is of the form: 
\begin{equation}
 D_{\rm K} \approx 3 \times 10^{27} \left( \frac {p}{mc} \right)^{1/3}  \eta \left( \frac{B}{3 \mu{\text{G}}} \right)^{-1/3} \left( \frac{L_{\rm c}}{50 \text{pc}} \right)^{2/3} \text{cm}^2 \,\text{s}^{-1},
\end{equation}
where $L_{\rm c}$ is the coherence length of the Galactic turbulent magnetic field. This functional form is derived from applying quasi-linear theory to a Kolmogorov spectrum with $(\delta B/B)^{2}=1$. However, in order to mimic the acceleration at oblique shocks \citep{jokipii}, we introduce a parameter $\eta<1$, which corresponds to smaller diffusion coefficient and higher maximum momentum of electrons,  whenever the age of the remnant is the limiting factor for acceleration. We adopt $L=20$ pc, $B_1=3$ $\mu$G and $\eta=0.01$ as reference values, keeping in mind that larger values of $\eta$ or $L$ lead to smaller maximum momenta. 
For the three considered cases, $\tau_{\rm acc}$, $\tau_{\rm sync}$ and $t_{\rm age}$ are shown in Fig.~\ref{fig:times},  with values of the parameters that are appropriate for SN 1006 ($u_1=4.3 \times 10^8$cm $s^{-1}$). The maximum momentum of electrons is calculated through Eq.~\ref{eq:tacc_min}, and is $\sim 7 \times 10^{4}$ GeV, $\sim 3 \times 10^{4}$ GeV,  and $ \sim 7 \times 10^{2}$ GeV for cases (1), (2) and (3) respectively. 

In the following section, we use these estimates to compute the  spectrum of reaccelerated electrons at the shock of SN 1006 and the subsequent gamma--ray emission. 

\section{The case of SN 1006} 
\label{sec:SN1006}
SN 1006 is one of the best studied SNRs. It is the remnant of a supernova observed in the year 1006, and is famous for being the first detected SNR emitting non--thermal hard X--rays, possible evidence of efficient particle acceleration~\citep{koyama1995}. Since then, it has been detected by several instruments including Chandra, XMM--Newton, H.E.S.S. and more recently \textit{Fermi}--LAT.  Some of these observations have revealed a peculiar asymmetrical morphology, with two  bright regions in the north--east (NE) and south--west (SW) parts of the SNR. In the X--ray domain, observations of non--thermal X--ray filaments with Chandra and XMM--Newton have revealed a small--scale structure in the NE region~\citep{bamba2003,berezhko2003,long2003,li2018} indicating the presence of large magnetic fields (of order $\sim 100 \;  \mu$G) and an orientation of the background magnetic field in the NE--SW direction. In the very--high--energy range, observations with H.E.S.S. have confirmed the bilateral morphology~\citep{acero2010}. 
Theoretical work has led to the proposal that such morphology may be due to the structure of the background magnetic field, with a gamma--ray emission enhanced in the regions where the shock is quasi--parallel~\citep{volk2003,caprioli2017} and particle acceleration may proceed efficiently. 

Recent observations performed by Fermi--LAT have reported on the detection of gamma rays from the NE rim, but intriguingly not from SW~\citep{condon2017}. Attempts to explain this asymmetry in the GeV range have been formulated in terms of a recent interaction of the shock with a dense cloud~\citep{dubner2002,miceli2014,miceli2016}. 
The bilateral morphology is an argument in favor of a type Ia progenitor and typical values for the shock velocity have been derived for both regions in the range $3 - 7.3 \times 10^{8}$ cm/s~\citep[see][and discussions therein]{winkler2014}, and we adopt as reference value $u_1=4.3 \times 10^8$ cm/s as in~\cite[]{morlino2010}. 
The average ISM density has been estimated to be in the range $0.05-0.1$ cm$^{-3}$, and we adopt as a reference value $n_1=0.05$ cm$^{-3}$~\citep{acero2007}. Finally, the radius of the SNR shell has been estimated  to be $R_{\rm sh}\approx 7.7$ pc for a distance $1.8-2.0$ kpc~\citep{acero2010}. 

In order to determine the contribution of reaccelerated electrons and protons to the overall gamma--ray flux of SN 1006, we assume for $f_{\infty}^p$ and $f_{\infty}^e$ the LIS spectrum as presented in Sec.~\ref{sec:spectrum} and we calculate the spectrum of reaccelerated protons and electrons with the reference values of the parameters listed above.  Gamma rays from ICS are calculated following~\citet{blumenthalgould1970}, who calculated the production rate of gamma rays in the scattering of relativistic electrons on a seed photon field. Here we rely on the reformulation proposed by~\citet{khangulyan2014}, and implemented in the Naima package~\citep{naima}. Gamma rays from pion decay are calculated following~\citet{kafexhiu2014}. To estimate the contribution from the entire SNR, we consider a volume $V= \xi \; 4 \pi R_{\rm sh}^2 \times L$. The length $L$ is the width of the region containing reaccelerated particles and is estimated as $L \approx u_2\times  \text{min} (t_{\rm age}, \tau_{\rm sync})$. In addition, in order to account for complex morphology (such as the bilateral appearance of SN~1006), we introduce a filling factor $\xi$ and assume in the following a reference value $\xi=0.5$. For the leptonic emission, due to ICS, the photon backgrounds that we include are the optical, near--infrared and far--infrared, with typical temperatures $2.72$ K, $30$ K and 3000 K respectively and corresponding energy densities 0.261, 0.5 and 1 eV cm$^{-3}$.

\subsection{Gamma rays from DSRA}

\begin{figure}
\includegraphics[width=.5\textwidth]{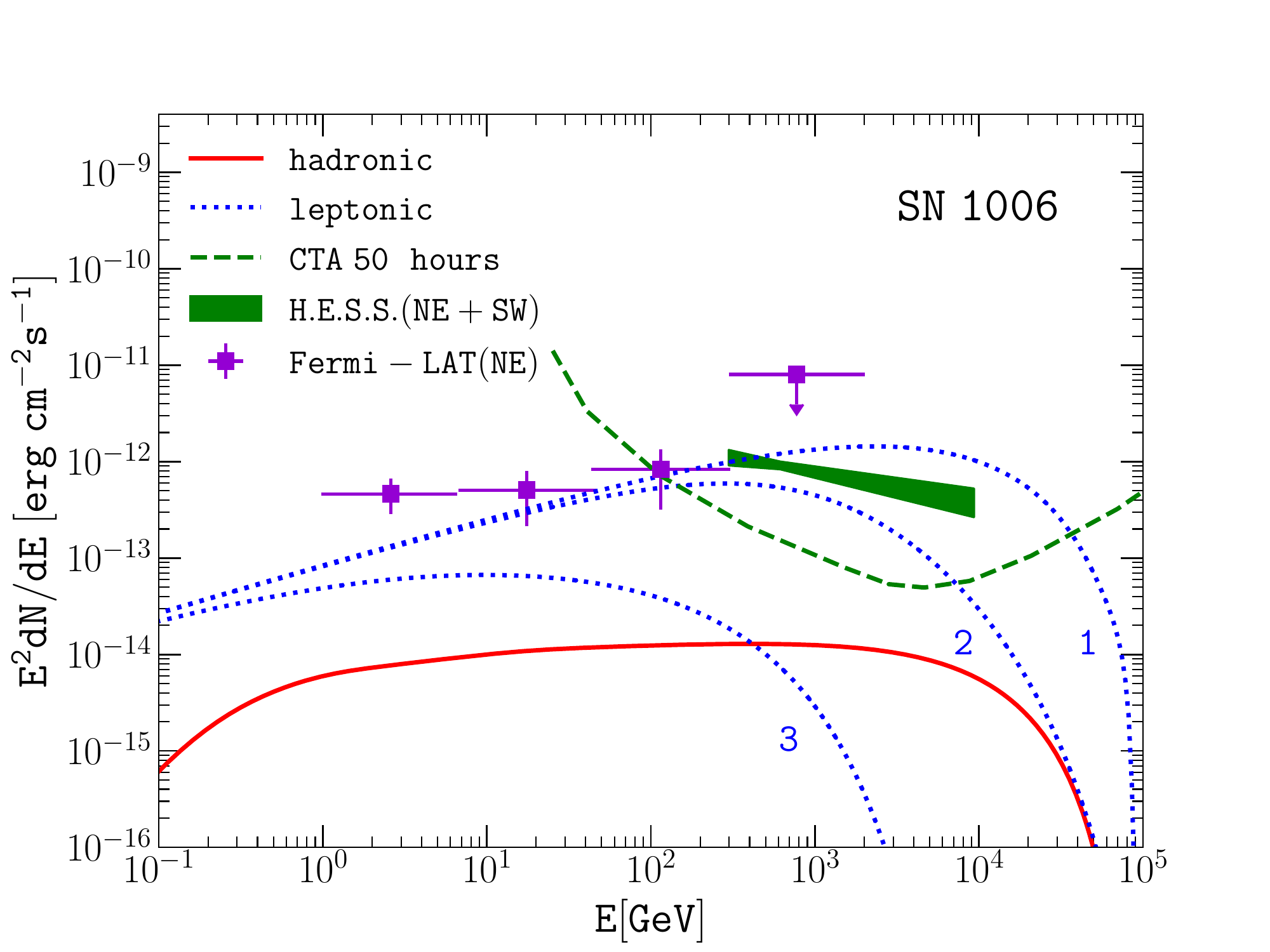}
\caption{Gamma rays from reaccelerated electrons (blue dotted line) and protons (red solid line), for the entire SNR. Three situations are considered for the electrons: (1) Bohm regime with $B_2= r \; B_1$ and $B_1= 3\;  \mu$G, (2) Bohm regime with $B_2=100 \; \mu$G, and (3) Kolmogorov regime, as discussed in the text. Observations of the NE region by Fermi--LAT (violet squares) and of the NE and SW regions by H.E.S.S. (green shade) are displayed. The sensitivity of CTA corresponds to a point source observed during 50 hours (green dashed line)~\citep{CTA}.}
\label{fig:SN1006_1}
\end{figure}

The gamma--ray emission from hadronic interactions (red solid line) and from leptonic interactions (blue dotted line) are shown in Fig.~\ref{fig:SN1006_1}. In Sec.~\ref{sec:spectrum}, we discussed the importance of the magnetic field and their impact on the maximum momentum of reaccelerated particles. 
The comparison with the measurements from Fermi--LAT (NE region) and H.E.S.S. (NE and SW regions), shown as violet squares and as a green area respectively, illustrates that the leptonic contribution obtained from reaccelerated electrons for typical values of the parameters describing the environment of the SNR ($n_1=0.05$ cm$^{-3}$, $d=2$ kpc, $R_{\rm sh}=7.7$ pc, and $u_1=4.3 \times 10^8$ cm s$^{-1}$) may be very close to current measurements. The gamma--ray component from hadronic interactions is subdominant. This is not surprising, given the low density of the ISM and that the hadronic component scales as $\propto n_1$.

As we discussed above, we consider three cases for particle diffusion at the shock. In case (1) we assume that the magnetic field downstream is $B_2= r \; B_1= 12 \; \mu$G and diffusion is Bohm--like. This situation minimizes the upstream magnetic field and leads to a loss-dominated maximum energy  for electrons $p_{\rm max}^e \sim 70$ TeV. Such a low value of the downstream magnetic field would not be compatible with the thin X-ray filaments observed from the NE and SW regions, and would better apply to the dark sides of the remnant, assuming that even in such regions diffusion is Bohm-like. On the other hand, our calculations show that the gamma--ray emission for half remnant as due to reaccelerated electrons is by itself of the same order of magnitude as the emission from the  gamma--ray bright regions in the NE and SW. Therefore, the current non--detection of some part of the SN 1006 already provides us with a strong constraint on the conditions for reacceleration in the two regions that have not been detected in gamma rays. We discuss this point further below. 

In the second diffusion model, case (2), we assume a value for the field downstream of the shock $B_2=100 \; \mu$G. Such value has been derived from measurement of X--ray filaments in the NE region~\citep{bamba2003,volk2003}, but not in the gamma--ray dark regions. We still consider this $B_2$ value, as an example, and illustrate how the resulting spectrum is affected, with a cut--off momentum $p_{\rm max}^e \sim 28$ TeV (due to more severe energy losses). Remarkably, the gamma--ray emission due to reaccelerated electrons, even in this case, saturates the gamma--ray emission in the TeV energy region and future CTA observations should provide better and more spatially resolved information about this emission. 

As discussed above, in order to suppress the gamma--ray emission due to reaccelerated electrons, one would need to adopt a substantially larger diffusion coefficient. This is plausible if reacceleration in the two gamma--ray dark sides of the remnant are considered. Our case (3) of diffusion may be useful to mimic this situation: a Kolmogorov regime is considered with a magnetic field downstream $B_2= 12 \; \mu$G, leading to a maximum momentum $p_{\rm max}^{e} \sim 700$ GeV, dominated by the age of the remnant. In this regime, the increased acceleration time $\tau_{\rm acc}$ leads to a significantly smaller value of the maximum momentum as compared to cases (1) and (2) (Bohm regime), as illustrated in Fig.~\ref{fig:times}. As a consequence, the gamma--ray emission from ICS of reaccelerated electrons, shown in Fig.~\ref{fig:SN1006_1} as case (3), falls below the data and eventually starts dropping at energy $\gtrsim 10$ GeV.  

We can then draw the following conclusions about SN1006: 1) for diffusion models that are appropriate to the quasi-parallel regions of the shock (Bohm-like), the ICS emission from reaccelerated electrons is of the same order of magnitude and with the same spectral shape as shown by the data in the GeV-TeV band. This contribution is weakly model dependent, in that the spectrum and normalization of the reaccelerated electrons are fixed once the LIS flux of electrons is measured. This fact might be interpreted as a detection of the reaccelerated electrons in SN1006, or as a tool to constrain the acceleration of fresh electrons at the shock. 2) The contribution of reaccelerated protons is subdominant in terms of gamma--ray emission. This is mainly due to the low density around SN1006. 3) The main uncertainty in assessing the role of reaccelerated electrons in the two dark sides of SN1006 is in the diffusion coefficient to be adopted there. If in such regions the diffusion coefficient at the shock is the same as deduced from secondary--to--primary ratios in the Galaxy, then the maximum energy of electrons is too low to imply any appreciable gamma--ray emission. On the other hand, we know that there is radio emission detected from such regions, which implies that electrons of at least $\gtrsim$ few GeV must be present (radio emission at 1.4 GHz is measured). Moreover, from the theoretical point of view, it has been speculated that in regions where the shock is oblique, acceleration might proceed faster \cite[]{jokipii}. We mimic this situation by assuming that the Galactic-like diffusion coefficient is reduced by some amount, so as to have maximum energy of electrons in the $\sim 1$~TeV range. In this case, as shown in Fig. ~\ref{fig:SN1006_1}, indeed reaccelerated electrons are insufficient to generate appreciable gamma--ray emission. For highly oblique shock configurations the maximum energy could be higher and the gamma--ray emission might be detectable. If future CTA observations will not reveal any appreciable gamma--ray emission from the two dark sides of SN1006, then we would have to conclude that the obliquity of the shock is inconsequential in terms of particle energization. 
Fig.~\ref{fig:SN1006_1} displays the expected sensitivity for CTA in the case of a point source observed for 50 hours (green dashed line), illustrating the possibility for CTA to detect the entire SNR shell because of reaccelerated electrons. Since SN1006 will appear as an extended source, the sensitivity is likely to be somewhat reduced but is in any case expected to be significantly improved compared to current Cherenkov instruments~\citep{CTA}.

%

\subsection{Comparison with DSA}

As discussed in the previous section, the gamma--ray emission measured from SN1006 is basically saturated by the ICS contribution of reaccelerated electrons, calculated assuming that this remnant is immersed in the same CR background as the solar system is. This appears to be a realistic assumption since the gradients in the proton and electron CR spectra are not prominent for energies $\lesssim 100$ GeV. This raises some questions about the acceleration of fresh particles at the shock (DSA), in the regions where it is expected to be quasi-parallel. 

In order to estimate the contribution from DSA, we assume that a faction $\xi_{\rm CR}=0.1$ of the shock ram pressure $\rho_1 u_1^2$ is converted into protons, and that the spectrum of DSA protons follows $f_p(p) \propto p^{-4}$. We assume that the electrons from DSA follow the proton spectrum with a  renormalization factor $K_{\rm ep}$ that for illustrating purposes is assumed to be $10^{-3}$, keeping in mind that this value is often estimated to be in the range $10^{-2}-10^{-4}$.  This approach is of course a simplified version of DSA at a SNR shock, not taking into account all deviations from the $-4$ power--law index expected when taking into account non--linear effects, losses (in the case of electrons), or particle escape. Nevertheless such simplified model is sufficient for the comparison we intend to make here. 

\begin{figure}
\includegraphics[width=.5\textwidth]{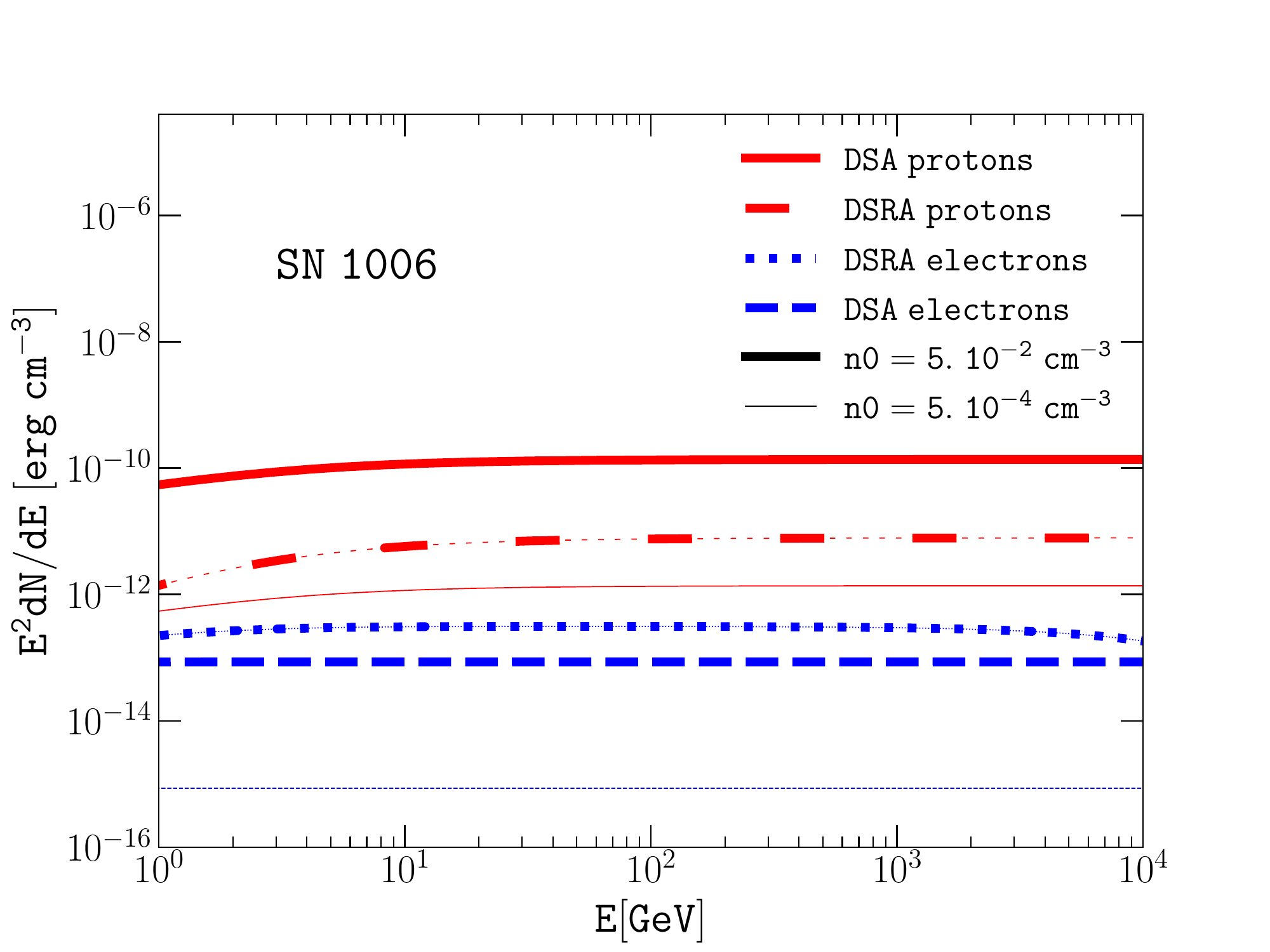}
\caption{Particle acceleration at SN 1006. Protons from DSA  and DSRA are shown as red solid and red short dashed lines, respectively. Electrons from DSA and DSRA correspond to blue dashed and blue dotted lines, respectively.  The thick and thin lines refer to densities $n_1=5 \times 10^{-2}$ cm$^{-3}$ and $n_1=5 \times 10^{-4}$ cm$^{-3}$.}
\label{fig:Particles_low_density}
\end{figure}

Fig.~\ref{fig:Particles_low_density} shows the spectrum of protons and electrons from DSA and DSRA for $n_1=5 \times 10^{-2}$ cm$^{-3}$ (thick lines), the estimated gas density in the surroundings of SN1006. Protons and electrons from DSA correspond to horizontal solid red and dotted blue lines, while protons and electrons from DSRA are shown as red short dashed and blue dashed lines. For DSRA, we assume that the proton and electron spectra far upstream of the shock are the same as in the LIS. Model (2) is adopted for the diffusion coefficient. 

One can see that the absolute normalization of the proton spectrum from DSA is dominant upon that of protons from DSRA, as one could naively expect. On the other hand, the situation is different for electrons: electrons from DSA and from DSRA are basically at the same level, and the spectral shape of the two components is also the same. This is simply due to the fact that DSRA is independent upon gas density while the spectrum of electrons accelerated through DSA has a normalization proportional to the gas density and to $K_{\rm ep}$. It follows that, for a given value of $K_{\rm ep}$, if the density is sufficiently low, the contribution of DSRA is dominant. In order to illustrate this simple fact, we consider a speculative situation in which the gas density is lowered to $n_1=5 \times 10^{-4}$ cm$^{-3}$ (thin lines). One can clearly see that in this case the contribution of electrons from DSRA is much larger than that of DSA. For this extreme choice of the gas density, even protons from DSA are subdominant compared to protons from DSRA. 

It may be useful to notice that when a small gas density is assumed close to a SNR shock, then even DSRA may induce non-linear effects as the pressure carried by the reaccelerated particles becomes close to $\rho_{1} u_{1}^{2}$. Such non--linear effects have not been taken into account here. 



\section{Other supernova remnants}
\label{sec:others}

From the discussion above, it is clear that the effects of DSRA are expected to be more prominent for SNRs that explode in underdense media. These SNRs are also expected to have a gamma--ray emission due to leptonic processes rather than hadronic interactions, since the latter scale linearly with gas density. Here we consider the case of two well studied SNRs detected in the gamma--ray domain: RX J1713--3946 and RX J0852.0--4622 (Vela Jr), detected by Fermi--LAT and H.E.S.S.~\citep{RXJFermi,VelaFermi,RXJHESS,VelaHESS}. For both remnants we investigate the expected contribution due to DSRA.
 
Previous studies of RX J1713--3946 have led to an estimated average ISM density $n_1= 0.02$ cm$^{-3}$, a distance $d \approx 1$~kpc, a radius $R_{\rm sh}=10$ pc and a magnetic field downstream of the order $B_2 \sim 100 \; \mu$G~\citep[see e.g.][and reference therein]{fukui2003,RXJHESS}. These numbers should be taken with much caution, in that they are derived from different indicators. For instance, the gas density relevant for gamma--ray production might be much higher than the average if there are dense clumps of material in the medium where the shock is expanding (see for instance \cite[]{gabici2014,fukui2017,Celli}). The magnetic field might be in the $0.1-1$ mG range in selected regions while being smaller in others, depending on where such field is the result of CR driven instabilities or rather fluid instabilities. 

\begin{figure}
\includegraphics[width=.5\textwidth]{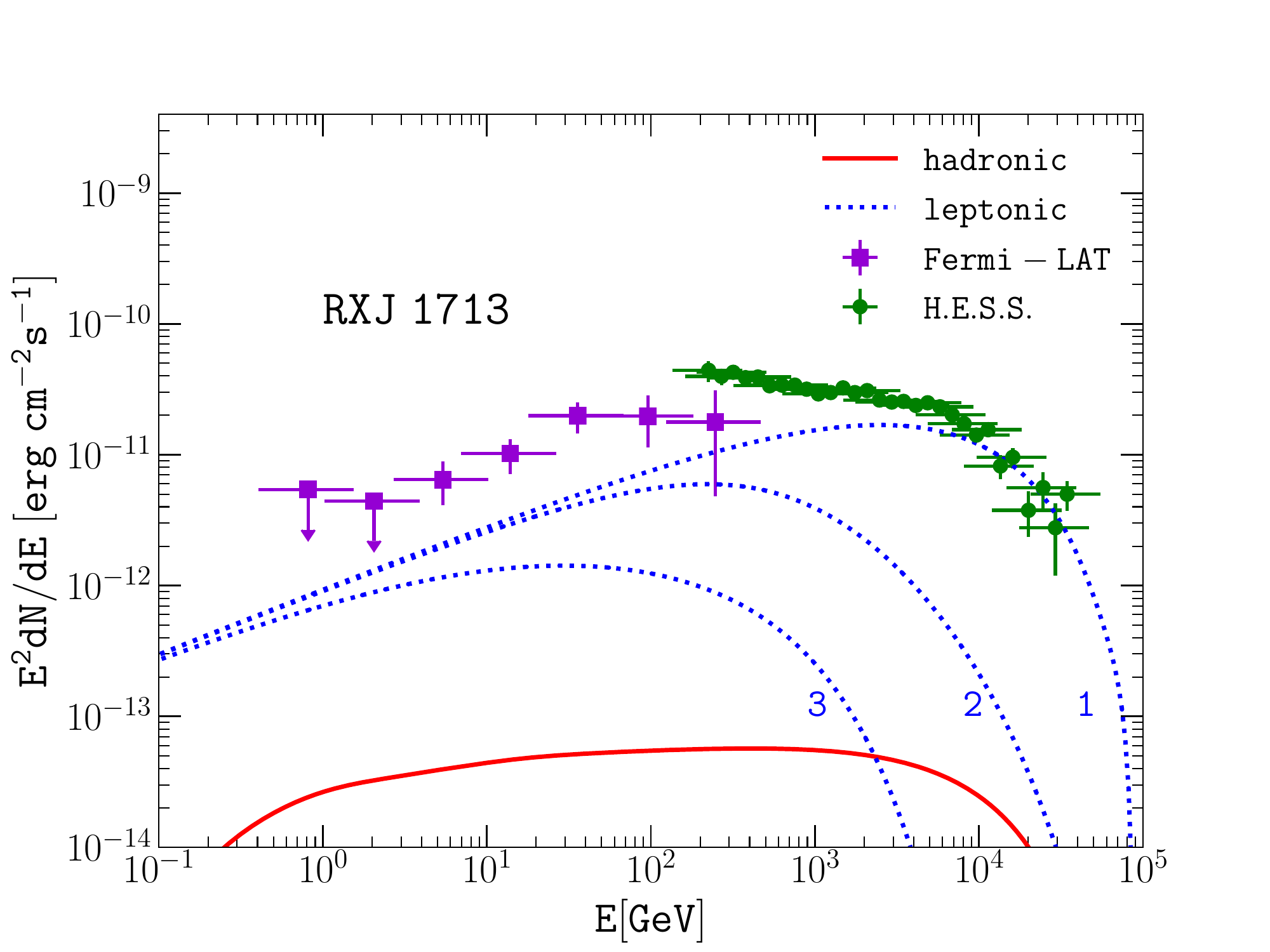}
\caption{Gamma rays from reaccelerated electrons (blue dotted line) and protons (red solid line) at RXJ1713--3946. As in Fig.~\ref{fig:SN1006_1}, three situations are considered for the electrons: (1) Bohm regime with $B_2= r \; B_1$ and $B_1= 3 \; \mu$G, (2) Bohm regime with $B_2=100 \; \mu$G, and (3) Kolmogorov regime with $B_1= 3 \; mu$G. 
Observations of the entire SNR by Fermi--LAT (violet squares) and by H.E.S.S. (green shade) are shown.}
\label{fig:RXJ}
\end{figure}

The gamma--ray spectrum from RX J1713--3946 as produced by reaccelerated particles is shown in Fig.~\ref{fig:RXJ}. As for SN 1006, three situations are considered for the electrons: (1) Bohm regime with $B_2= r \; B_1$ and $B_1= 3 \; \mu$G, (2) Bohm regime with $B_2=100 \; \mu$G, and (3) Kolmogorov regime with $B_1= 3 \; \mu$G. We adopted a conservative approach here and decided to retain a value $\xi=0.5$ for the volume filling factor of the regions where DSRA occurs with a given diffusion model. 
 
The differential spectrum measured for the entire SNR by Fermi--LAT and H.E.S.S. is shown. The normalization of the flux of gamma rays due to DSRA is slightly smaller than the observed flux for cases (1) and (2). For case (3), since the maximum energy of reaccelerated electrons is very low, the corresponding gamma--ray emission is too low to describe observations. Cases (1) and (2), or a case that is intermediate between the two, would lead to approximately the correct maximum energy of electrons to describe H.E.S.S. observations. The normalization of the gamma--ray flux from reaccelerated particles also deserves some comments: first, as noticed above, we retained the same value $\xi=0.5$ adopted for SN1006, but this does not need to be so. Second, the gamma--ray emission from ICS may well be slightly larger if the photon background close to the remnant is higher.  

Once again, the observed flux of gamma rays is remarkably close to what one would expect if the spectrum of electrons, assumed to be the same as in the LIS, get reaccelerated at the shock, with no unambiguous need for freshly accelerated particles. Notice that this does not imply that efficient particle acceleration is not taking place, in that the low gas density might prevent the gamma--ray signal to be dominated by pion decays. On the other hand, if models based on the clumpy structure of the ISM in the vicinity of the remnant could find an independent confirmation, one would be left with the non trivial job of finding a reason for reacceleration to be subdominant. 

\begin{figure}
\includegraphics[width=.5\textwidth]{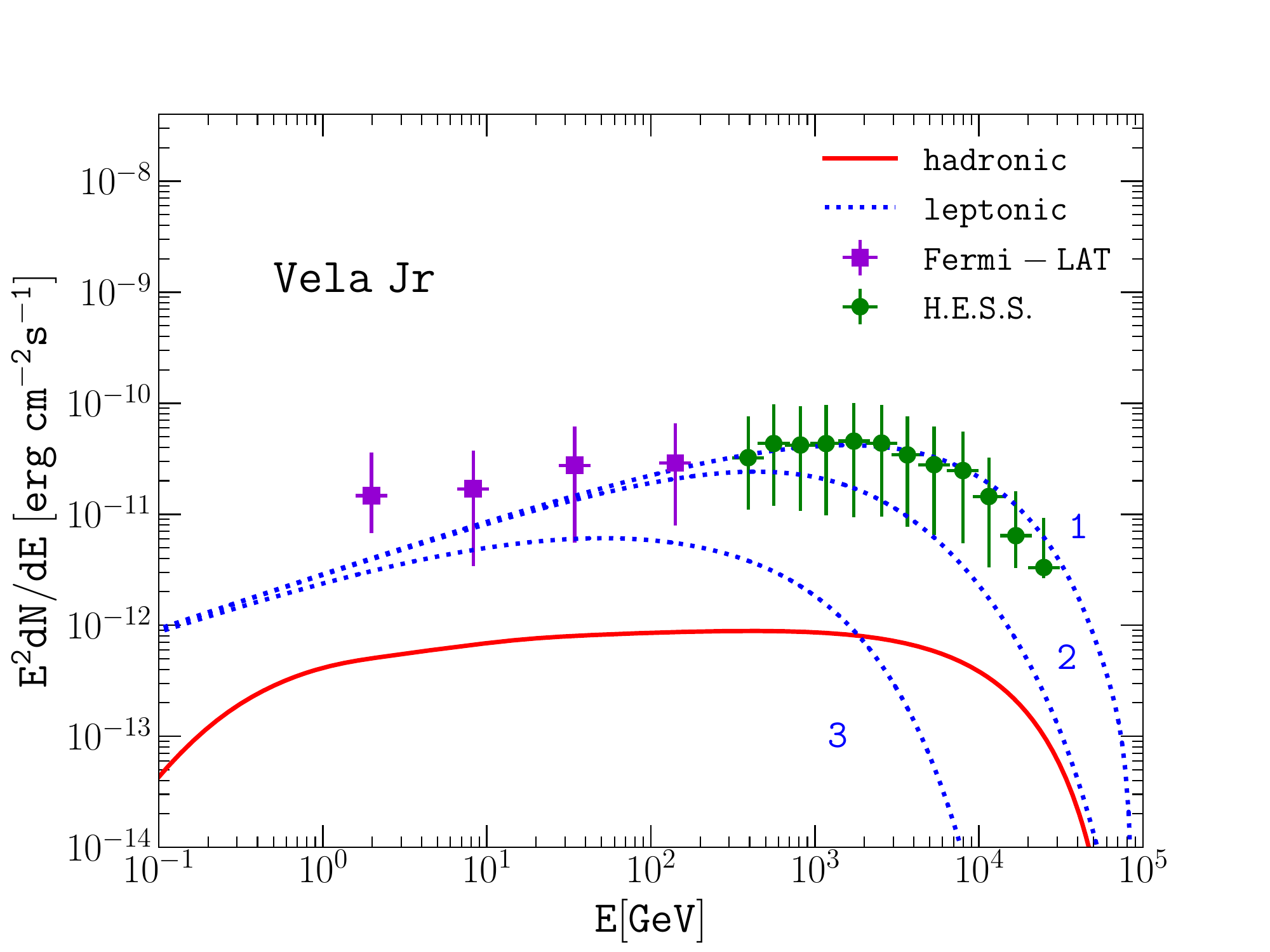}
\caption{Gamma rays from reaccelerated electrons (blue dotted line) and protons (red solid line) at Vela Jr. 
As in Fig.~\ref{fig:SN1006_1}, three situations are considered for the electrons: (1) Bohm regime with $B_2= r \; B_1$ and $B_1= 3$ $\mu$G, (2) Bohm regime with $B_2=40$ $\mu$G, and (3) Kolmogorov regime. 
Observations of the entire SNR by Fermi--LAT (violet squares) and by H.E.S.S. (green shade) are shown.}
\label{fig:vela}
\end{figure}

The case of Vela Jr is shown in Fig.~\ref{fig:vela}. The properties of this very well studied SNR are comparable to those of RX J1713--3946. We rely on current estimates of the parameters and adopt $u_1=3\times 10^8$cm s$^{-1}$, $d = 0.750$~kpc, $R_{\rm sh}=11.8$~pc, $n_1=0.1$~cm$^{-3}$ and a magnetic field downstream of the shock  $B_2=40\; \mu$G~\citep{katsuda2008,allen2015,VelaHESS}.  As for the previous calculations, the gamma--ray luminosity from DSRA electrons is computed for the three diffusion models in the shock region. 

Very similar considerations to those made above for SN1006 and RX J1713--3946 also apply to Vela Jr, which strengthens the case for a gamma--ray emission from these remnants that may be dominated by ICS of reaccelerated electrons. On the other hand, this finding makes the unambiguous identification of sources of freshly accelerated protons even more problematic and possibly limited to SNRs located near dense regions. Unfortunately most of such cases are typically middle age SNRs for which the time of efficient acceleration is past\footnote {Interestingly, even for such remnants the case has been made of reacceleration of CR hadrons \cite[]{w44}.}. One possible exception to this rule is the Tycho SNR, for which the estimated density is high enough to make a solid case in favor of efficient acceleration and gamma--ray emission of hadronic origin~\cite[]{MorlinoTycho,BlasiTycho}.

\section{Conclusions}
\label{sec:concl}

Gamma--ray observations of SNRs are the most powerful tool at our disposal to make a solid case in favor or against SNRs as the main source of the bulk of CRs in our Galaxy. Yet, at present, such observations have opened more problems that they settled: 1) many of the sources that have been detected have hard gamma--ray spectra, often compatible with the ICS emission from accelerated electrons; 2) in a handful of cases, gamma--ray spectra steeper than $E^{-2}$ have been measured, incompatible with ICS of electrons, and hinting at a hadronic explanation. However such observations suggest a CR spectrum that is appreciably steeper than naively expected from DSA, and even more at odds with the non-linear extensions of DSA \cite[]{Caprioli2011}. Moreover in all these cases the gamma--ray spectra did not show evidence for proton acceleration to energies above $\sim 100$ TeV, well away from the knee region. 3) In middle aged SNRs the observed gamma--ray emission can certainly be described by hadronic interactions with pion decays. However these old remnants are past their optimal age as CR accelerators and typically their spectra are steep, and in fact it has been proposed that the gamma--ray emission can be due to interactions of reaccelerated CR hadrons \cite[]{w44}, rather than freshly accelerated particles. 

In addition to all this, the reacceleration of particles at SNR shocks has recently been investigated as a possible mechanism responsible for some anomalous behaviour observed in secondary CR nuclei and antiprotons \cite[]{blasi2017,bresci2019}. Moreover, the first {\it ab initio} simulations of particle acceleration in the presence of seeds have recently been performed, including the possible excitation of waves through streaming of reaccelerated particles \cite[]{caprioli2017}. 

This situation triggered our interest in looking for additional implications of DSRA, perhaps on the gamma--ray emission of individual SNRs. Here we studied three SNRs, SN 1006, RX J1713--3946 and Vela Jr., all similar in that the inferred value of the surrounding gas density is relatively low. This condition makes it easier to identify the contribution of particles reaccelerated through DSRA. 

Energization of seed particles is a quite interesting process: provided the energy of the seeds is higher than the injection energy, a condition that is easily satisfied, they all get accelerated at a shock, with a spectrum that reflects the strength of the shock. For a strong shock, the final spectrum is very similar to $p^{-4}$ unless the spectrum of seeds is harder than $p^{-4}$, in which case it is left unchanged. The energy channelled into the reaccelerated particles only depends on the compression factor at the shock and on the maximum energy that can be reached. The latter can be estimated following the same lines of thought as for ordinary DSA. This implies that not much model dependence appears once the spectrum of the seeds is fixed. 

In our cases, the seed spectrum is that of protons and electrons measured in the LIS, for which we have Voyager I \cite[]{cummings2016} PAMELA \cite[]{adriani2011,adriani2011b} and AMS-02~\citep{aguilar2015,aguilar2019} measurements and careful parametrizations of such data~\citep{bisschoff2019}. The main source of uncertainty in calculating the spectra of electrons and protons reaccelerated at a SNR shock is the determination of the diffusion coefficient in the shock proximity, since it defines the acceleration time and the maximum energy. For this reason, we adopted three phenomenological recipes for the diffusion coefficient, so as to bracket the range of possible situations. 

These diffusion models have been used for the calculation of the spectra of protons and electrons in SN 1006, RX J1713--3946 and Vela Jr. and the corresponding gamma--ray emission. Remarkably, for all three SNRs the gamma--ray flux from GeV to TeV can easily be accommodated in terms of ICS of CR electrons reaccelerated at SNR shocks, with no clear indication of the presence of freshly accelerated particles. The main constraint comes from SN1006, which shows a bilateral morphology: in the bright lobes, the magnetic field is inferred from X-ray measurements and the X-ray cutoff is well described if the diffusion coefficient is Bohm-like. Within these assumptions, electrons from DSRA are sufficient to account for the gamma--ray emission. In other words, in such regions, there is little room to accommodate electron acceleration from the thermal plasma ($K_{\rm ep}\lesssim 10^{-3}$). In the two lobes of SN1006 that are bright in gamma rays, DSRA can only be switched off by assuming that the diffusion coefficient is similar to that measured in the Galaxy, perhaps suppressed by effects related to shock obiquity. This situation leads to exceedingly low maximum energy of reaccelerated electrons. Therefore, observations of the two dark lobes in the gamma--ray domain can help constrain the diffusion coefficient in the vicinity of SN~1006. 

For the other two SNRs that have been considered here, the measured flux from the whole remnant can be compared with the predicted one from DRSA of electrons. We stress again that the flux and spectrum of such electrons only depends upon the compression factor of the shock (not its velocity or local density) and the scattering properties. Again, the observations are easily matched to the ICS emission of reaccelerated electrons, with no clear need for freshly accelerated electrons (or protons). If any, the only indirect proof that efficient particle acceleration of protons is taking place is the presence of large magnetic fields, that reflect in a filamentary X-ray morphology and may be due to the excitation of CR streaming instabilities upstream of the shock, at the locations where the shock is quasi-parallel.

\section*{Acknowledgment}
PB and PC thank C. Evoli and G. Morlino for useful discussions. The authors acknowledge use of the Python package Naima~\citep{naima}. The research of PB was partially funded through Grant ASI-INAF n. 2017-14-H.0.






\bsp	
\label{lastpage}
\end{document}